\documentclass[fleqn,usenatbib]{mnras}
\usepackage{newtxtext,newtxmath}
\usepackage[T1]{fontenc}
\usepackage{graphicx}	
\usepackage{amsmath}	
\usepackage{xcolor}
\usepackage{threeparttable}
\usepackage{orcidlink}

\DeclareRobustCommand{\VAN}[3]{#2}
\let\VANthebibliography\thebibliography
\def\thebibliography{\DeclareRobustCommand{\VAN}[3]{##3}\VANthebibliography}

\title[LTT 9779 b with HST/WFC3 UVIS]{Constraining the Scattered Light properties of LTT 9779 b Using HST/WFC3 UVIS}

\author[Radica et al.]{
Michael Radica$^{\orcidlink{0000-0002-3328-1203}~1,2}$\thanks{E-mail: radicamc@uchicago.edu}\thanks{NSERC Postdoctoral Fellow},
Jake Taylor$^{\orcidlink{0000-0003-4844-9838}~3}$,
Hannah R.\ Wakeford$^{\orcidlink{0000-0003-4328-3867}~4}$,
David Lafrenière$^{\orcidlink{0000-0002-6780-4252}~2}$,
\newauthor
Romain Allart$^{\orcidlink{0000-0002-1199-9759}~2}$\thanks{SNSF Postdoctoral Fellow},
Nicolas B.\ Cowan$^{\orcidlink{0000-0001-6129-5699}~5,6}$,
James S. Jenkins$^{\orcidlink{0000-0003-2733-8725}~7,8}$,
and Vivien Parmentier$^{\orcidlink{0000-0001-9521-6258}~9}$
\\
$^{1}$Department of Astronomy \& Astrophysics, University of Chicago, 5640 South Ellis Avenue, Chicago, IL 60637, USA\\
$^{2}$Trottier Institute for Research on Exoplanets, Université de Montréal, 1375 Avenue Thérèse-Lavoie-Roux, Montréal, QC, H2V 0B3, Canada\\
$^{3}$Department of Physics, University of Oxford, Parks Rd, Oxford, OX1 3PU, UK\\
$^{4}$School of Physics, University of Bristol, H.H.~Wills Physics Laboratory, Tyndall Avenue, Bristol BS8 1TL, UK\\
$^{5}$Department of Physics, McGill University, 3600 rue University, Montréal, QC, H3A 2T8, Canada\\
$^{6}$Department of Earth and Planetary Sciences, McGill University, 3600 rue University, Montréal, QC, H3A 2T8, Canada\\
$^7$Instituto de Estudios Astrof\'isicos, Facultad de Ingenier\'ia y Ciencias, Universidad Diego Portales, Av. Ej\'ercito 441, Santiago, Chile\\
$^8$Centro de Astrof\'isica y Tecnolog\'ias Afines (CATA), Casilla 36-D, Santiago, Chile\\
$^9$Laboratoire Lagrange, Observatoire de la Côte d’Azur, Université Côte d’Azur, Nice, France
}

\date{Accepted XXX. Received YYY; in original form ZZZ}
\pubyear{\the\year{}}

\begin{document}

\label{firstpage}
\pagerange{\pageref{firstpage}--\pageref{lastpage}}
\maketitle

\begin{abstract}
A planet's albedo is a fundamental property that sets its energy budget by dictating the fraction of incident radiation absorbed versus reflected back to space. Generally, optical eclipse observations have revealed the majority of hot, giant planets to have low albedos, indicating dayside atmospheres dominated by absorption instead of reflection. However, there are several exceptions to this rule, including the ultra-hot-Neptune LTT 9779\,b, which have been found to have high geometric albedos. We observed four eclipses of LTT 9779\,b with the G280 grism of the Hubble Space Telescope's WFC3 UVIS mode; targeting the scattering signatures of the cloud condensate species causing the planet's elevated reflectivity. However, we do not definitively detect the planet's eclipse in our observations, with injection-recovery tests yielding a 3-$\sigma$ upper limit of 113\,ppm on the eclipse depth of LTT 9779\,b in the 0.2--0.8\,µm waveband. We create reflectance spectrum grids for LTT 9779b's dayside using \texttt{VIRGA}/\texttt{PICASO} and compare to our UVIS limit, as well as previously published CHEOPS and TESS eclipse photometry. We find that silicate condensates are best able to explain LTT 9779\,b's highly-reflective dayside. Our forward model grids only enable weak constraints on vertical mixing efficiency, and suggest that, regardless of their particular composition, the clouds are likely composed of smaller and more reflective particles. Our work facilitates a deeper understanding of the reflectance properties of LTT 9779\,b as well as the UVIS spectroscopic mode itself, which will remain the community's primary access to UV wavelengths until next-generation telescopes like the Habitable Worlds Observatory. 
\end{abstract}

\begin{keywords}
planets and satellites: atmospheres -- planets and satellites: gaseous planets -- planets and satellites: individual: LTT 9779 b
\end{keywords}

\section{Introduction} 
\label{sec: Introduction}

Measurements of secondary eclipses of hot, giant exoplanets at infrared wavelengths have proven to be excellent probes of the planet's thermal emission \citep[e.g.,][]{charbonneau_detection_2005, deming_infrared_2005, seager_dayside_2005, deming_spitzer_2015, beatty_evidence_2017, evans_ultrahot_2017, arcangeli_h_2018, baxter_transition_2020, dragomir_spitzer_2020, mansfield_unique_2021, changeat_five_2022, coulombe_broadband_2023}. However, secondary eclipse measurements at optical wavelengths also open the possibility of observing light reflected from the planet's atmosphere \citep[e.g.,][]{sudarsky_albedo_2000, seager_photometric_2000, demory_inference_2013, evans_deep_2013, angerhausen_comprehensive_2015, fraine_dark_2021, brandeker_cheops_2022, hoyer_extremely_2023, krenn_geometric_2023, taylor_another_2023}. 

Observations with the Hubble Space Telescope (HST) Space Telescope Imaging Spectrograph (STIS) as well as with photometric instruments such as the Transiting Exoplanet Survey Satellite (TESS), the Characterising Exoplanets Satellite (CHEOPS), and \textit{Kepler} have shown that the daysides of highly-irradiated planets are generally dark with albedos of $\rm A_g \lesssim 0.3$ \citep[e.g.,][]{evans_deep_2013, angerhausen_comprehensive_2015, esteves_changing_2015, bell_very_2017, fraine_dark_2021, Wong2021, brandeker_cheops_2022, krenn_geometric_2023, singh_cheops_2024}. Particularly for ultra-hot (T$\rm_{eq}>2000$\,K) planets, these measurements have generally been interpreted to mean that their daysides are systematically too hot for significant condensation of clouds to occur \citep{parmentier_thermal_2018, mansfield_unique_2021}. Theoretical models suggest that their optical spectra are instead dominated by absorption due to alkali metals such as Na and K, and potentially metal oxides like TiO and VO, instead of reflection from high-albedo clouds \citep{sudarsky_albedo_2000, seager_photometric_2000}.

\subsection{LTT 9779 b: A Highly-Reflective Ultra-Hot-Neptune}

There are, however, several exceptions to this general trend including Kepler-13\,Ab\footnote{though the high inferred albedo of the ultra-hot Kepler-13\,A b is likely due to residual thermal emission at optical wavelengths instead of reflected light.} \citep{shporer_atmospheric_2014, Wong2021}, Kepler-7\,b \citep{demory_inference_2013}, and LTT 9779\,b \citep{dragomir_spitzer_2020, hoyer_extremely_2023}, each of which has been found to have an elevated albedo via optical eclipse measurements, with LTT 9779\,b, in particular, presenting an intriguing puzzle. LTT 9779\,b is a $\rm 29.32 \pm 0.8\,M_\oplus$ and $\rm 4.72 \pm 0.23\,R_\oplus$ planet, which with an orbital period of 0.792\,d resides right in the middle of the hot-Neptune desert \citep{szabo_short-period_2011, mazeh_dearth_2016, jenkins_ultrahot_2020}. Observations with Spitzer's Infrared Array Camera (IRAC) by \citet{dragomir_spitzer_2020} and \citet{crossfield_phase_2020} revealed that despite residing in such an inhospitable region of the exoplanet parameter space, LTT 9779\,b still retains a H/He-dominated atmosphere --- a finding corroborated by \citet{edwards_characterizing_2023} with HST. More recently, \citet{radica_muted_2024}, \citet{vissapragada_high-resolution_2024}, and \citet{reyes_closer_2025} used transmission observations at both low, and high spectral resolution, to demonstrate that LTT 9779\,b's atmosphere metallicity is likely extremely elevated, with \citet{radica_muted_2024} estimating a range of 20--850$\times$ solar. \citet{radica_muted_2024} also found evidence for high-altitude clouds in the planet's atmospheric terminator. Despite having a dayside temperature of $\sim$2300\,K \citep{dragomir_spitzer_2020} which places it firmly in the regime of ultra-hot planets, \citet{hoyer_extremely_2023} found a geometric albedo of $\rm A_g \sim 0.8$ using CHEOPS eclipse observations. Moreover, they argued that only in atmospheric metallicities in excess of 400$\times$ solar could clouds form high enough in the planet's atmosphere to explain the observed albedo. 

\citet{coulombe_highly_2025} then presented an optical-to-near-infrared phase curve of LTT 9779\,b observed with the Single Object Slitless Spectroscopy (SOSS; 0.6--2.85\,µm) of JWST's NIRISS instrument \citep{albert_near_2023, doyon_near_2023}. They find a lower albedo than \citet{hoyer_extremely_2023} of $\rm A_g = 0.5 \pm 0.07$ --- though one that is still significantly elevated relative to the wider population of hot, giant planets. Moreover, with the phase-resolved spectral information gained from the phase curve, they find that the dayside reflectivity is highly-asymmetric, with the western dayside having an albedo of $\rm A_g = 0.79 \pm 0.15$ and a much lower $\rm A_g = 0.41 \pm 0.1$ on the eastern dayside --- consistent with the circulation of high-albedo clouds from the nightside onto the dayside via the western terminator \citep[e.g.,][]{parmentier_thermal_2018, parmentier_cloudy_2020}. 

\subsection{HST/UVIS: A New Avenue to Probe Exoplanet Reflectivity}

Particularly for ultra-hot planets, both thermal and reflected light can contribute to the measured eclipse depth in optical wavebands, which can complicate the inference of a geometric albedo \citep{heng_understanding_2013, schwartz_balancing_2015, parmentier_transitions_2016}. A potential solution to this issue is to move to blue-optical, or even near-ultraviolet (NUV) wavelengths where the flux contributions from the planet's own thermal emission, even for ultra-hot planets, becomes negligible. \citet{fraine_dark_2021} made the first attempt to probe reflected light with the UVIS mode of HST's Wide Field Camera 3 (WFC3) by observing a single eclipse of WASP-43\,b with the UVIS F350LP photometric filter (0.35--0.82\,µm). They did not detect an eclipse from the planet --- indicative of a dark dayside hemisphere with $A_g \lesssim 0.07$.

UVIS though, is not only a photometric channel, but also has spectroscopic capabilities. \citet{wakeford_into_2020} and \citet{lewis_into_2020} observed two transits of the hot-Jupiter HAT-P-41\,b with the G280 grism (0.2--0.8\,µm) and conclude that UVIS can outperform STIS in terms of spectroscopic precision, while extending the accessible wavelengths down to 0.2\,µm (compared to 0.3\,µm for STIS). Even more recently, \citet{lothringer_uv_2022} and \citet{boehm_hustle_2025} used the G280 grism to observe transits of WASP-178\,b and WASP-127\,b respectively. However, to date no eclipse observations have been published with the UVIS G280 grism. 

G280 eclipse observations not only provide the opportunity to calculate a dayside albedo via a broadband eclipse measurement, but the spectroscopic capabilities also offer the possibility of definitively detecting the particular cloud condensates causing the elevated albedo via their scattering signatures --- particularly at wavelengths $<$0.6\,µm \citep[e.g.,][]{sudarsky_albedo_2000}. The absorption features of individual cloud condensates have been detected in mid-infrared transmission and emission spectroscopic observations \citep{grant_jwst-tst_2023, dyrek_so2_2024, inglis_quartz_2024}. Optical and ultraviolet transit studies, both from the ground and from space, have only yielded evidence for scattering slopes \citep[e.g.,][]{pont_detection_2008, sing_continuum_2016, kirk_rayleigh_2017, gao_aerosols_2021, taylor_awesome_2023}, whereas eclipse observations have lacked the precision to effectively probe the unique scattering signatures of individual cloud condensate species.   

In this paper, we attempt the first spectroscopic eclipse study of an exoplanet with the HST/WFC3 UVIS G280 grism targeting LTT 9779\,b. This paper is organized as follows: Section~\ref{sec: Observations} describes the observations and data reduction. We outline our cloud model grid results in Sections~\ref{sec: Models}. We then discuss these results in Section~\ref{sec: Discussion} before concluding in Section~\ref{sec: Conclusions}.

\section{Observations and Data Analysis} 
\label{sec: Observations}

We observed four eclipses of LTT 9779\,b using HST's WFC3 UVIS mode (GO 16915, PI: Radica). Using the G280 grism provides spectroscopic wavelength coverage from 0.2 -- 0.8\,µm. The four visits occurred on 2022 Jun 05, Jun 06, Jun 07, and Aug 07, with each visit consisting of three HST orbits; one before, one during, and one after eclipse. We used 60\,s exposures resulting in a total of 60 exposures per visit. As in \citet{wakeford_into_2020}, we use a 2250$\times$650 subarray and $-$50" POSTARG Y-offset such that the spectrum is centered on chip 2 and we capture both the +1 and $-$1 spectral orders of the target. We additionally specify strict orientation requirements to avoid contamination from two nearby stars. 

\begin{figure} 
	\centering
	\includegraphics[width=\columnwidth]{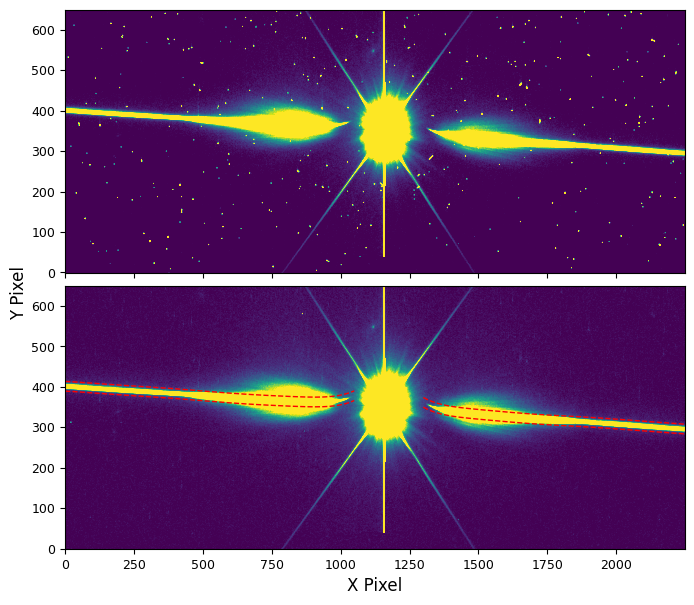}
    \caption{Example HST/WFC3 UVIS G280 data frame before (top) and after (bottom) processing, shown on a linear colour scale. The bottom panel has been bias and background subtracted, as well as cleaned of cosmic ray hits. The aperture around the +1 (left) and $-$1 (right) spectral orders used for the extraction are denoted in red in the bottom panel. 
    \label{fig:UVIS data}}
\end{figure}

UVIS observations share many similarities with those using JWST NIRISS/SOSS (e.g., slitless observations, curved spectral traces, overlapping orders). As such, we reduced the UVIS exposures with a custom version of the \texttt{exoTEDRF} pipeline \citep{feinstein_early_2023, radica_awesome_2023, radica_exotedrf_2024, radica_promise_2025}, tweaked to handle HST data. For each exposure, we start with the \texttt{\_flt.fits} files produced by the \texttt{calwfc3} pipeline. We first subtract a custom bias frame which was taken at the end of each visit --- though we note that the use of this bias correction had negligible impact on the final results. We then apply the \texttt{exoTEDRF} cosmic ray detection algorithm \citep{radica_awesome_2023, radica_muted_2024} which flags and replaces pixels which deviate by more than 5$\sigma$ from a running median in time with that median. As in \citet{wakeford_into_2020} and \citet{boehm_hustle_2025}, we follow this by a spatial cosmic ray cleaning using Laplacian Edge Detection \citep{vandokkum_cosmicray_2001}.

To estimate the background level in each frame, we used the median value of all non-illuminated pixels ($>$50 pixels away from the spectra trace). We found the median to be a more robust estimation of the background level than the mode \citep[e.g.,][]{wakeford_into_2020, boehm_hustle_2025}, which displayed multiple, seemingly spurious, increases or decreases in the background level from exposure to exposure. Moreover, we do not see the peaks in the background level at the start of each new HST orbit reported by \citet{wakeford_into_2020}. An example of a final, cleaned data frame compared to its original, unprocessed, state is shown in Figure~\ref{fig:UVIS data}.

We locate the positions of both the +1 and $-$1 order traces using a modified version of the \texttt{edgetrigger} algorithm \citep{radica_applesoss_2022}. We then extract the spectra by summing all flux within a box aperture with a half-width of 12 pixels, which was found to minimize the light curve scatter. As with NIRISS/SOSS observations, the +2 and -2 orders partially overlap the +1 and $-$1 order traces at wavelengths longer than $\sim$0.4\,µm \citep{wakeford_into_2020}. In these observations, the second order traces reach a maximum brightness of only $\lesssim$1\% of the first order traces, and we thus ignore this contamination for the remainder of our analysis\footnote{This level of contamination could potentially be important in a full spectroscopic analysis, however our non-detection of an eclipse at the white light level precluded this.} Finally, we perform the wavelength calibration of our extracted spectra using the equations from \citet{pirzkal2017} \citep[and updates from][]{pirzkal2020} and our extracted trace positions.

\subsection{Light Curve Fitting}
\label{sec: Light Curves HST}

We fit the light curves using a ``jitter decorrelation" approach \citep[e.g.,][]{sing_hubble_2019, wakeford_into_2020}, whereby we detrend against outputs of HST's Pointing Control System. We consider a total of 14 jitter vectors: $t$, a linear time trend, $\phi_{HST}^i$, HST's orbital phase for which we consider up to fourth-order polynomials (i.e., $i<=4$), $\delta_\lambda$, the wavelength shift of the spectrum on the detector, $\delta_x$ and $\delta_y$, the x and y positions of the PSF on the detector, $V2\_roll$ and $V3\_roll$, the roll angle of the telescope along the V2 and V3 axes, $LAT$ and $LONG$, the aperture latitude and longitude, and finally $RA$ and $DEC$, the aperture right ascension and declination. Most parameters are retrieved from the \texttt{\_jit.fits} files, except for the wavelength and PSF pixel shifts which we calculate from the data as follows: for $\delta_\lambda$, we cross correlate the extracted spectra of the +1 and $-$1 orders with a PHOENIX stellar template \citep{husser_new_2013} matching the physical parameters of LTT 9779, and find shifts on the order of 35--40\,\AA\ as a function of exposure. For $\delta_x$ and $\delta_y$, we cross correlate the 2D detector image with a median stack of all exposures in a given visit.

For each of the four visits, we use the \texttt{juliet} framework \citep{espinoza_juliet_2019} to jointly fit a \texttt{batman} \citep{kreidberg_batman_2015} eclipse model and the systematics model consisting of the jitter vectors outlined above to white light curves of each order. A benefit of the \texttt{juliet} framework is that fitting can be carried out using \texttt{pymultinest} \citep{buchner_statistical_2016}, a byproduct of which is a direct calculation of the Bayesian evidence. This means that we can use Bayesian model comparison to select the optimal combination of jitter vectors. We thus test all combinations of our 14 jitter vectors to find the optimal ensemble for each order and each visit. 

\begin{figure} 
	\centering
	\includegraphics[width=\columnwidth]{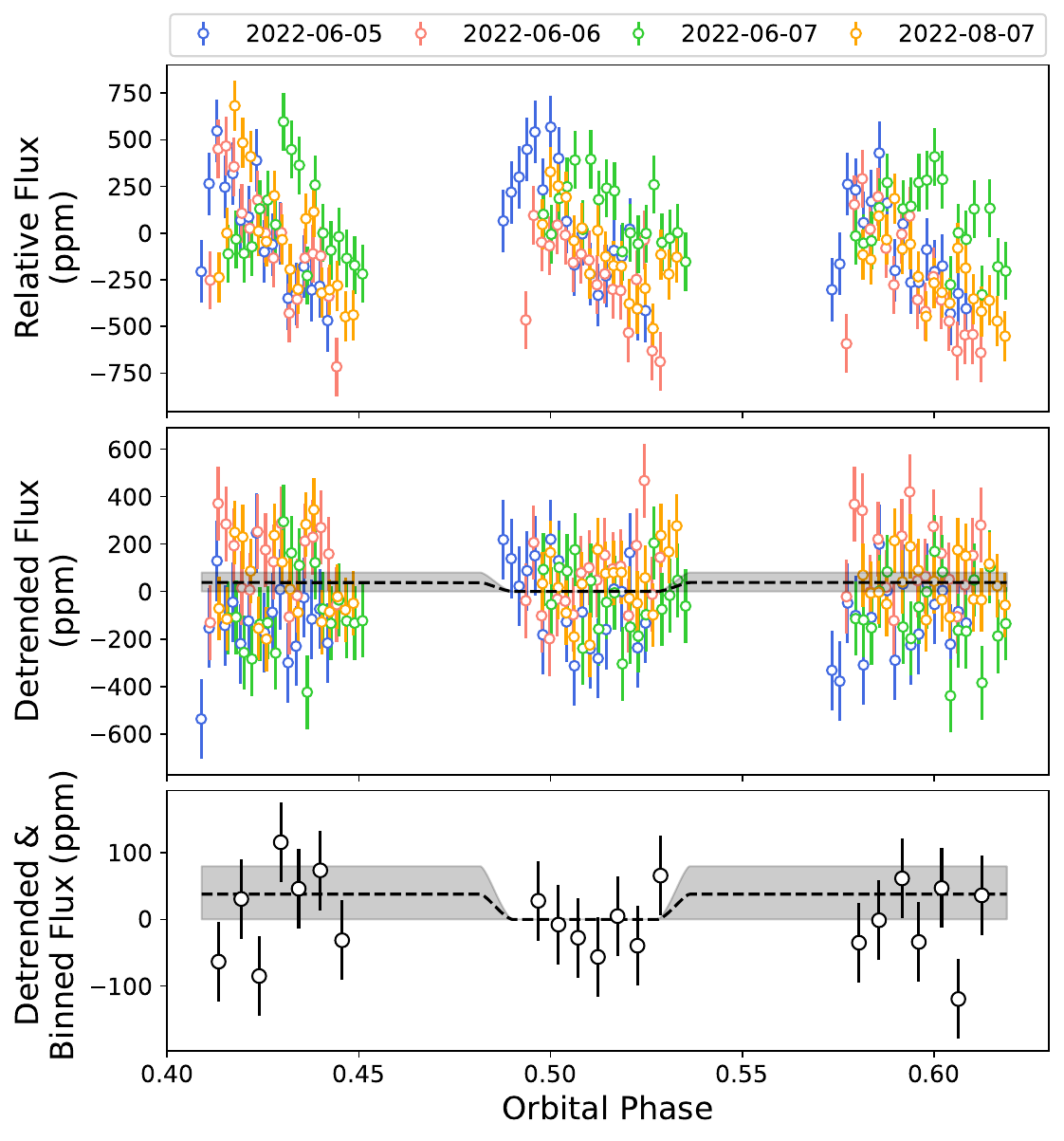}
    \caption{Phase-folded HST WFC3 UVIS/G280 white light curves (0.2 -- 0.8\,µm) for all four eclipses. 
    \emph{Top}: Raw white light curves, created through a weighted average of all flux in the +1 and $-$1 orders. Points are coloured according to the observation date. 
    \emph{Middle}: White light curves after removing the optimal systematics model from each visit. The best-fitting eclipse model is shown in the black solid line, and the grey shading denotes the 1.5$\sigma$ credible envelope. 
    \emph{Bottom:} Same as middle, except the observations are binned in increments of 10 exposures for visualization purposes.
    \label{fig:UVIS wlc}}
\end{figure}

As shown in the top panel of Figure~\ref{fig:UVIS wlc}, there is no clear sign of an eclipse in the white light curves of any visit (even when stacking both orders). Moreover, the observations have virtually no coverage of the eclipse ingress or egress, which are necessary to obtain robust constraints on the planet's orbital parameters \citep{seager_unique_2003}. Therefore, we fix all orbital parameters (i.e, the mid-eclipse time, orbital eccentricity, inclination, and semi-major axis) to the best fitting values from \citet{coulombe_highly_2025}, who presented a full phase curve of LTT 9779\,b observed with JWST NIRISS/SOSS (GTO 1201, PI: Lafrenière) starting on Jul 7, 2022 --- between the third and fourth visits of this UVIS program. Our only free astrophysical parameters are thus the eclipse depth, $F_p/F_*$, and the eclipse zero point which compensates for any errors in the baseline normalization. We use wide, uninformative priors for both parameters and allow for negative eclipse depths. 

We find no definitive evidence for an eclipse in either order of any visit, likely because the residual scatter is still too large in any single white light curve alone. Since the +1 and $-$1 orders appear to show similar systematic trends, we also tried fitting a weighted average of the two orders for each visit, though this offered no improvement (Figure~\ref{fig:UVIS wlc}). Finally, we removed the best-fitting jitter decorrelation model from each combined white light curve, and simultaneously fit an eclipse model to all four visits, from which we obtain an eclipse depth of $F_p/F_* = 38.55 \pm 26.54$\,ppm --- consistent with a null result at the 1.5-$\sigma$ level (shown in black in the bottom panel of Figure~\ref{fig:UVIS wlc}). Moreover, via Bayesian model comparison we find that a flat line is preferred to our best-fitting eclipse by $\Delta \ln Z = 4.6$, or $\sim$3.5$\sigma$ by the \citet{benneke_how_2013} scale. We also verified that this is not a function of the data reduction techniques used, as we were able to replicate both the spectrum and precision of \citet{wakeford_into_2020} for HAT-P-41\,b using their observations. Nor is it a function of the light curve detrending as we find the same results using a systematics marginalization technique \citep[e.g.,][]{wakeford_marginalizing_2016, wakeford_into_2020}, or Gaussian process regression. Finally, we verified that our light curve residuals from each visit bin down as would be expected for white noise and performed a series of Kolmogorov-Smirnov \citep{massey_kolmogorov_1951} and Anderson-Darling \citep{anderson_asymptotic_1952} tests for normality, indicating that it is not the presence of significant amounts of residual correlated noise that is stymieing the detection of an eclipse in this data.\footnote{Summary of test results included in this \href{https://zenodo.org/records/14967486}{Zenodo} archive.} 

The throughput of WFC3/UVIS peaks at $\sim$0.25\,µm, and falls off towards longer wavelengths (see Fig.~1 in \citet{wakeford_into_2020}). It is therefore possible that our white light eclipses are biased to negligible values by a drop-off in reflectivity seen in certain cloud species at these wavelengths (e.g., MnS, CaTiO$_3$; \citealp{taylor_how_2021}). To test this hypothesis, we divide the 0.2--0.8\,µm UVIS waveband into three 0.2\,µm-wide bins and carry out the same fitting described above. However, we again find no evidence for an eclipse in any of our bins. 

We considered the possibility that we simply missed the planet's eclipse due to, e.g., using an incorrect, or out-of-date ephemeris for the HST observations. However, propagating the ephemeris used for the HST proposal (which was from the \textit{Spitzer} phase curve published by \citet{crossfield_phase_2020}) perfectly predicts the timing of the SOSS phase curve eclipses which we use to predict the eclipse time for our light curve fits. 

\subsubsection{Injection Recovery Tests}
\label{sec: Injection Recovery}

\begin{figure} 
	\centering
	\includegraphics[width=\columnwidth]{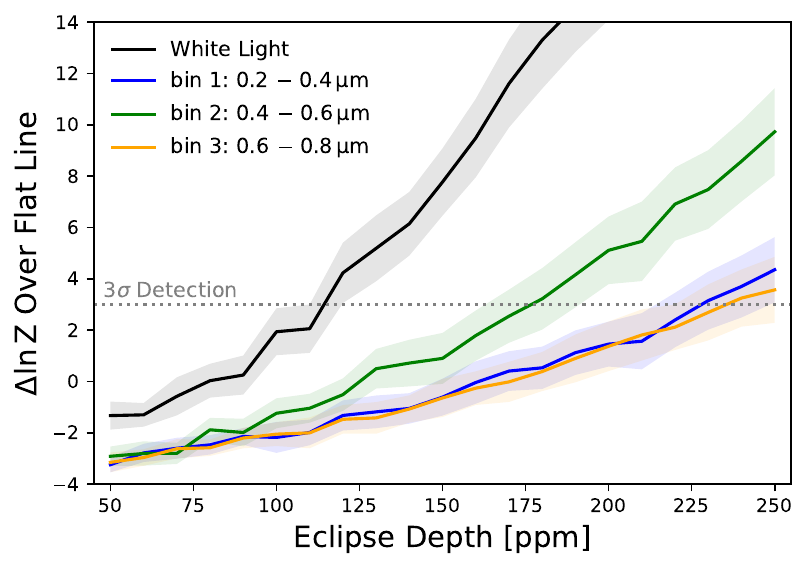}
    \caption{Flat line rejection significance (via a Bayesian evidence comparison) as a function of injected eclipse depth for the UVIS white light (black), as well as in three spectroscopic bins (blue, green, and orange). For each case, the solid lines are the median of 100 injection-recovery tests with the shaded regions showing the spread. The grey dotted line shows the threshold for a 3-$\sigma$ rejection of a flat line. This demonstrates that for our data quality we would have been able to robustly identify eclipses with depths greater than 113\,ppm, if present, thereby giving us an upper limit on the planets reflectivity. 
    \label{fig: Detection Significance}}
\end{figure}

In order to quantify the upper limit that we can place on the UVIS eclipse depth of LTT 9779\,b from our observations, we carry out a series of injection-recovery tests. We generate eclipse light curves, with the same time sampling as our observations for eclipses with depths spanning 50 -- 250\,ppm in increments of 10\,ppm. We then inject Gaussian noise with a standard deviation equal to the 155\,ppm scatter in the combined white light curve residuals. For each simulated observation, we attempt to retrieve the injected eclipse depth, following the same light curve fitting prescription described above. Finally, we repeat the same exercise but on a light curve with the same scatter and no injected eclipse. We show the detection significance (which we quantify as the difference in Bayesian evidence between the eclipse model and the flat line for a given simulation) as a function of injected eclipse depth as the black line in Figure~\ref{fig: Detection Significance}. We determine a 3$\sigma$ upper limit on the UVIS white light eclipse depth of LTT 9779\,b of 113\,ppm; almost identical to the eclipse depth reported by \citet{hoyer_extremely_2023} from observations with CHEOPS ($115 \pm 24$\,ppm).

We repeat the same procedure for each of the three 0.2\,µm wavelength bins described above. The detection significance curves are also shown in Figure~\ref{fig: Detection Significance} in the coloured lines. These yield 3-$\sigma$ upper limits of $\sim$180\,ppm for the 0.4--0.6\,µm bin and $\sim$230\,ppm for both the 0.2--0.4\,µm and 0.6--0.8\,µm bins. The lower upper limit in the middle bin is primarily driven by a combination of the UVIS throughput and the shape of the stellar spectrum, resulting in the extracted flux peaking at $\sim$0.5\,µm.

\section{Modelling the Reflectivity of LTT 9779 \lowercase{b}'s Dayside} 
\label{sec: Models}

\begin{figure*} 
	\centering
    \includegraphics[width=0.95\textwidth]{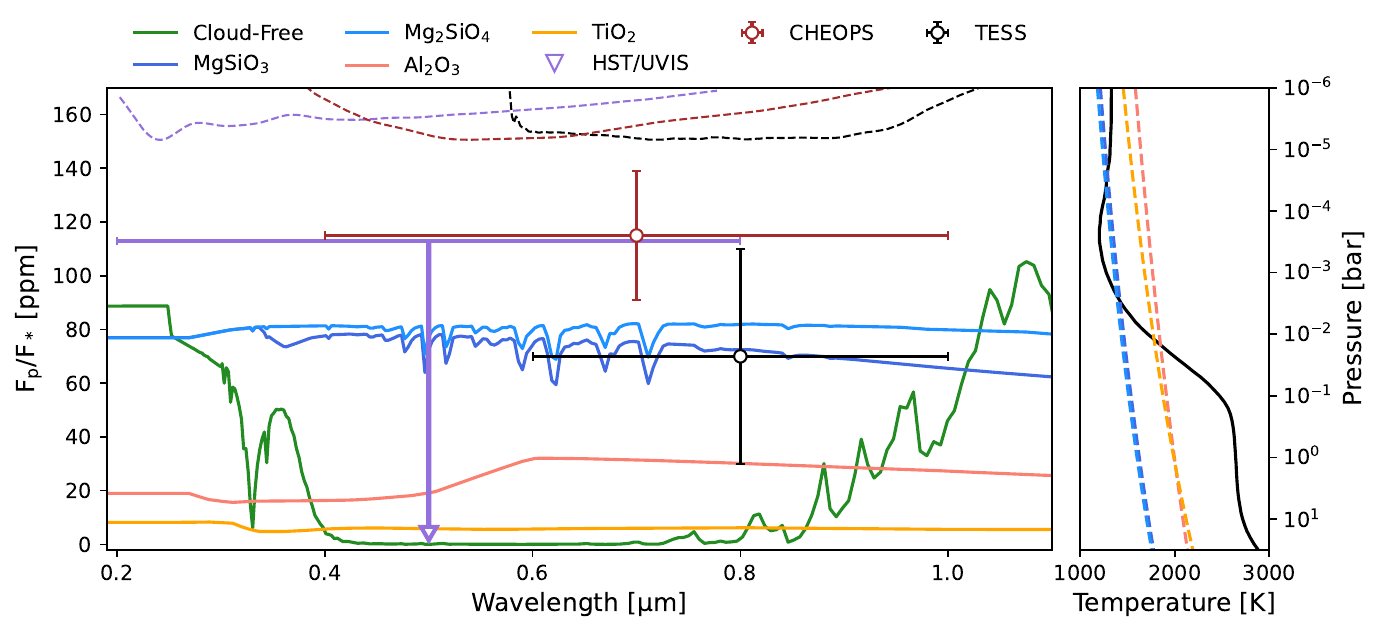}
    \caption{\emph{Left}: Comparison of UV/optical photometry with simulated reflectance spectra from the dayside atmosphere of LTT 9779\,b. The HST/WFC3 UVIS 3$\sigma$ upper limit derived in Section~\ref{sec: Injection Recovery} is denoted by the purple arrow, and we include previously published CHEOPS and TESS photometric eclipse depths in brown and black respectively. The throughputs for each instrument are shown in dashed lines at the top of the panel. Overplotted in the solid coloured lines are reflectance spectra generated with \texttt{VIRGA}/\texttt{PICASO} for several plausible cloud condensate species.
    \emph{Right}: Temperature-pressure profile assumed for the dayside atmosphere of LTT 9779\,b. Condensation curves from \citet{visscher_atmospheric_2010} for all considered species are overplotted in dashed lines.} 
    \label{fig: Spectrum}
\end{figure*}

In order to quantify the properties of LTT 9779\,b's atmosphere at optical/NUV wavelengths, we generate a grid of forward models using the open-source python packages \texttt{VIRGA} \citep{rooney_new_2022} to derive cloud properties, and \texttt{PICASO} \citep{batalha_exoplanet_2019} to compute reflectance spectra given the \texttt{VIRGA} outputs. Before assuming a cloud species, \texttt{VIRGA} will recommend (based on the condensation curves for various species) which cloud-condensate species are plausible, given an input atmospheric temperature-pressure (TP) profile and atmospheric metallicity. In its current configuration, \texttt{VIRGA} assumes the metallicity of the atmosphere is solar. For the TP profile, we generate a model consistent with that retrieved by \citet{coulombe_highly_2025}, from a retrieval on LTT 9779\,b's JWST NIRISS/SOSS dayside emission spectrum (Figure~\ref{fig: Spectrum}; right panel). Based on the above conditions, we find that the condensate species MgSiO$_3$, Mg$_2$SiO$_4$, Al$_2$O$_3$, and TiO$_2$ can potentially form at low pressures in the dayside atmosphere of LTT 9779\,b. 

We then use \texttt{PICASO} to generate emission spectra based on gas volume mixing ratios, our chosen TP profile, and assumed values for the cloud particle sedimentation efficiency, $f_\text{sed}$, and the strength of atmospheric vertical mixing parameterized by the eddy diffusion coefficient, $K_\text{zz}$, as outlined in \citet{ackerman_precipitating_2001}. We couple \texttt{PICASO} to \texttt{FastChem} to generate the gas volume mixing ratios assuming a metallicity of solar for self-consistency, and then build a grid of forward models for each cloud condensate species by varying the values of $f_\text{sed}$ and $K_\text{zz}$. We allow $f_\text{sed}$ to vary between 0.01 and 4, linearly in steps of 10, and $K_\text{zz}$ between 10$^5$ to 10$^{12}$, logarithmically in steps of 10. These values span the full domain of expectations for hot-Jupiters and solar system giant planets alike; from thick and extended, to thin and compressed clouds \citep{ackerman_precipitating_2001, parmentier_3d_2013, komacek_vertical_2019}.

For each cloud condensing species, we then perform a $\chi^2$-per-data-point ($\chi^2_d$) goodness-of-fit analysis comparing to optical observations of LTT 9779\,b. In addition to the HST/WFC3 UVIS upper limit derived in Section~\ref{sec: Injection Recovery}, we also include the CHEOPS and TESS eclipse depths from \citet{hoyer_extremely_2023} and \citet{dragomir_spitzer_2020} respectively. Since we are solely concerned here with the reflective properties of LTT 9779\,b's dayside atmosphere, we elect not to include the \textit{Spitzer} 3.6 and 4.5\,µm photometric nor JWST NIRISS/SOSS eclipse spectroscopic data in our analysis. These data are much more sensitive to the gas-phase composition of the atmosphere as opposed to the condensed phase, which preferentially impacts optical wavelengths through reflection, or MIR wavelengths via absorption \citep[e.g.,][]{grant_jwst-tst_2023, dyrek_so2_2024, inglis_quartz_2024}. Therefore, excluding them will have a minimal impact on the reflectance properties that we are able to derive. Moreover, we note that the optical portion of the NIRISS/SOSS eclipse spectrum is consistent with the TESS and CHEOPS depths, and that the TP profile we use is directly informed by these JWST observations.      

\begin{figure*} 
	\centering
	\includegraphics[width=0.95\textwidth]{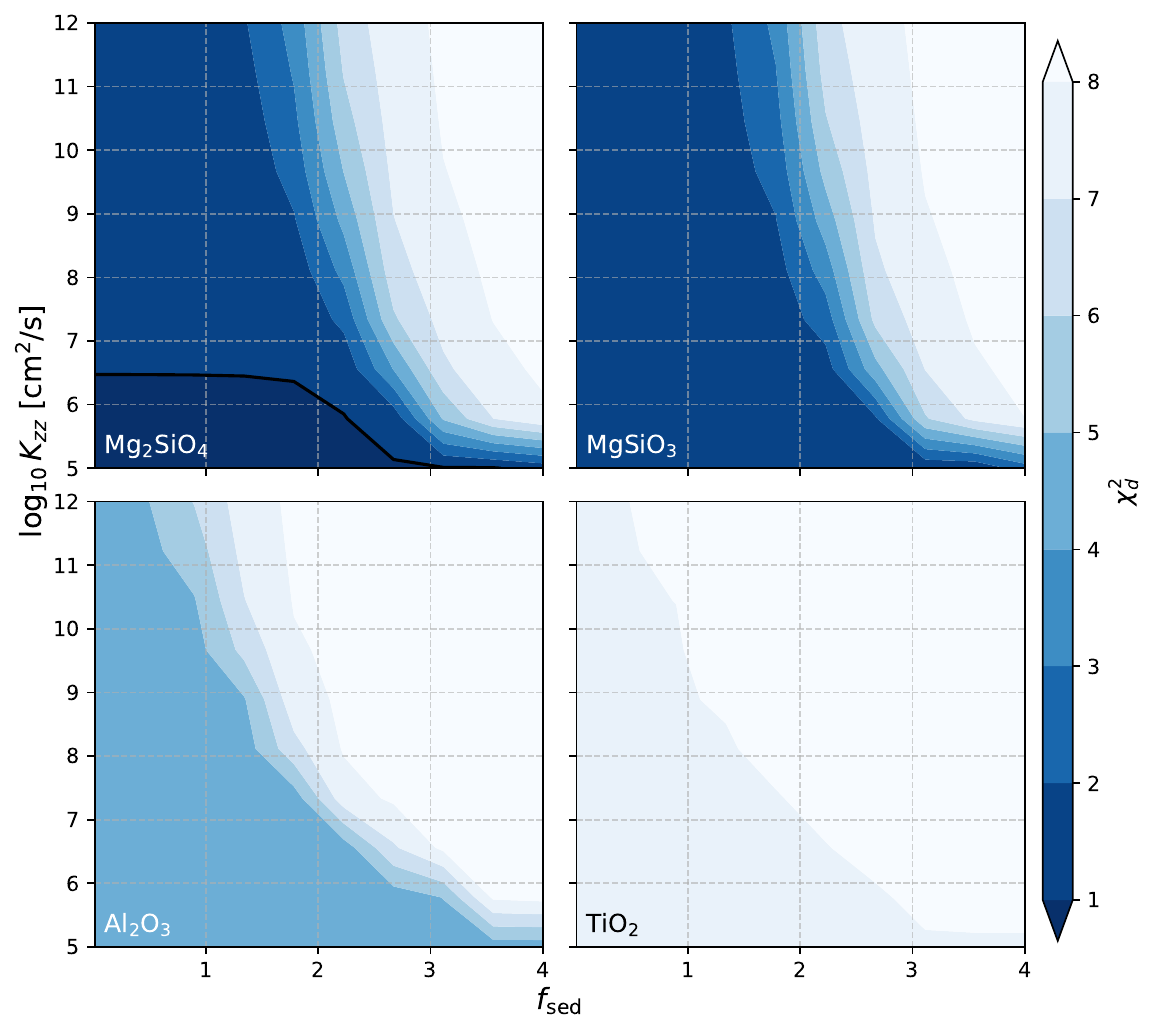}
    \caption{Goodness-of-fit maps in the $K_{zz}$ vs.\ $f\rm _{sed}$ parameter space for reflectance spectra calculated from four different cloud condensate species: MgSiO$_3$, Mg$_2$SiO$_4$, Al$_2$O$_3$, and TiO$_2$. Goodness-of-fit is evaluated using a $\chi^2$ per datum ($\chi^2_d$) metric, where smaller values (darker colours) indicate better fits. The solid black line in the Mg$_2$SiO$_4$ panel denotes $\chi^2_d$=1. Constraints on $K_{zz}$ are weak, but lower $f\rm _{sed}$ values are generally preferred, favouring small and highly-reflective condensate particles.   
    \label{fig: Chi2 Maps}}
\end{figure*}

We show the best-fitting reflectance spectra from each considered condensate species, along with the optical eclipse observations in Figure~\ref{fig: Spectrum}, and in Figure~\ref{fig: Chi2 Maps} we also show the resulting $\chi^2_d$ maps.

\section{Discussion} 
\label{sec: Discussion}

\subsection{Implications for LTT 9779 b's Atmosphere}

Figure~\ref{fig: Spectrum} demonstrates that, of the condensate species that we have considered in our analysis, only the silicates MgSiO$_3$ and Mg$_2$SiO$_4$ are able to adequately match the observations. Silicate clouds are one of the most common condensate species expected to form in the atmospheres of hot giant planets \citep{wakeford_transmission_2015, wakeford_high-temperature_2017, gao_aerosol_2020}, and this work adds to the growing body of literature that suggests that silicate clouds are causing the high dayside reflectivity observed for LTT 9779\,b. Indeed, \citet{hoyer_extremely_2023} and \citet{coulombe_highly_2025} came to similar conclusions through their analysis of the TESS, CHEOPS, and \textit{Spitzer} dayside photometry, as well as the NIRISS/SOSS emission spectrum respectively. 

Like \citet{hoyer_extremely_2023}, we find that Ti-based clouds, in our case TiO$_2$, are able to form in LTT 9779\,b's atmosphere. However, they never reach sufficient optical depth to be the sole cause for the planet's high observed dayside albedo. Though, as pointed out by \citet{hoyer_extremely_2023}, the formation of Ti-based condensates may remove sufficiently significant amounts of this strong optical absorber from the gas phase to stifle the formation of a thermal inversion \citep{hubeny_possible_2003, spiegel_can_2009}, as LTT 9779\,b has been confirmed to possess a non-inverted dayside temperature structure \citep{dragomir_spitzer_2020, coulombe_highly_2025} --- seemingly at odds with the broader population of ultra-hot planets \citep{baxter_transition_2020, mansfield_unique_2021, changeat_five_2022}. Al-based clouds, which are some of the highest-temperature condensates expected to form in the atmosphere of a hot giant planet \citet{wakeford_high-temperature_2017} and represented here by Al$_2$O$_3$ also do not reach sufficient optical depth to explain the observations.

Unlike both \citet{hoyer_extremely_2023} and \citet{coulombe_highly_2025}, who both tailor their cloud parameterizations to obtain the most reflective possible clouds, we perform a broader sweep of the parameter space of potential particle size distributions via the \citet{ackerman_precipitating_2001} $K_{zz}$/$f\rm_{sed}$ parameterization (also used in both of the above works). As shown in Figure~\ref{fig: Chi2 Maps}, irrespective of the particular condensate composition, our models prefer to reside in roughly the same region of parameter space. The constraints on the eddy diffusion strength, via $K_{zz}$, are weak in all cases, though lower values are more prevalent, particularly for Mg$_2$SiO$_4$. However, larger values more in line with common predictions from general circulation models for the atmospheres of hot Jupiters are not ruled out \citep{parmentier_3d_2013, komacek_vertical_2019, rooney_new_2022}. Similarly, lower values of $f\rm_{sed}$ are preferred in all cases, which directly leads to smaller, and therefore more reflective, cloud particles \citep{ackerman_precipitating_2001, gao_sedimentation_2018}. The results from our broad parameter sweep are therefore entirely consistent with the findings of both \citet{hoyer_extremely_2023} and \citet{coulombe_highly_2025}, despite our not tailoring our models \textit{a-priori} to increase reflectivity.

A potential caveat to our modelling analysis is that we have assumed that LTT 9779\,b's atmosphere is of solar composition in order to keep the gaseous chemistry self-consistent with the \texttt{VIRGA} cloud calculations. There is currently some uncertainty in the literature over the true metallicity of LTT 9779\,b's atmosphere. Based on mass-metallicity trends from the solar system and transiting exoplanets \citep[e.g.,][]{thorngren_massmetallicity_2016, welbanks_massmetallicity_2019} LTT 9779\,b would be expected to have a metallicity highly-enhanced relative to solar. Early work by \citet{crossfield_phase_2020} and \citet{hoyer_extremely_2023}, as well as analyses of the planet's JWST NIRISS/SOSS and ESPRESSO transmission spectra by \citet{radica_muted_2024} and \citet{reyes_closer_2025} respectively, support this conclusion, as does the lower-than-expected rate of atmosphere escape inferred by \citet{vissapragada_high-resolution_2024}. However, \citet{edwards_characterizing_2023} and \citet{coulombe_highly_2025} both find sub-solar metallicities from HST/WFC3 transit and JWST NIRISS/SOSS phase curve observations respectively. Upcoming JWST NIRSpec/G395H observations should provide a definitive measurement of the planet's atmosphere metallicity.

Our assumption of solar metallicity has an effect both on the gas-phase chemistry, as well as the types of condensates which can form. However, at the optical/NUV wavelengths that we are concerned with here, the gas-phase chemistry is only of secondary importance and would have a much greater effect in the NIR where prominent molecular bands of O- and C-bearing species are located. Increasing the atmosphere metallicity also broadly allows for condensates to form more readily at a given temperature \citep{wakeford_high-temperature_2017}. \citet{hoyer_extremely_2023} find that they require the atmospheric metallicity of LTT 9779\,b to be at least 400$\times$ solar before sufficient cloud formation occurs to match the measured CHEOPS albedo. However, our data-driven dayside TP profile, inferred from the NIRISS/SOSS eclipse observation, is generally cooler than their modelled profiles, which were based solely off of the planet's equilibrium temperature and a varying heat redistribution factor. Therefore, clouds are more readily able to form in our models without needing to invoke such high metallicities. Increasing the metallicity may allow for lower-temperature S- or Fe-based condensates to form \citep{wakeford_high-temperature_2017}, however Fe-based clouds are generally dark and would not lead to a high dayside albedo \citep{chubb_dark_2024}, whereas S-based condensates like ZnS or MnS have incredibly low nucleation rates and would be unlikely to form high-optical depth cloud layers \citep{gao_aerosol_2020}.

Finally we note that none of our modelled spectra are able to match the albedo inferred by the CHEOPS eclipse depth at 1-$\sigma$ (though they are still consistent at better than 2-$\sigma$). The CHEOPS depth is higher than both the TESS depth and UVIS upper limit, which bracket either side of the CHEOPS waveband. The upper limit inferred in the reddest UVIS band in Section~\ref{sec: Injection Recovery}, which significantly overlaps with the CHEOPS bandpass, is insufficiently precise to add any further constraint in this waveband. Therefore, we repeat our modelling analysis but remove the CHEOPS depth from consideration. The results are shown in Figure~\ref{fig: Chi2 Map No CHEOPS}, and are qualitatively similar to those presented above despite yielding overall better fits. LTT 9779\,b is set to be observed further with CHEOPS, and it will be instructive to see whether the new observations result in a revision of the eclipse depth to be more in line with the TESS and JWST observations.

\subsection{On the Performance of UVIS for Exoplanet Time Series Observations}

In the original observing proposal, we had intended to use HST/UVIS to create an emission spectrum of LTT 9779\,b from 0.2--0.8\,µm in order to identify the unique scattering signatures of individual cloud species. Using the UVIS simulator\footnote{\url{https://github.com/hrwakeford/HST_WFC3_UVIS_G280_sim}} developed as part of \citet{wakeford_into_2020}, and which has been shown to accurately reproduce the on-sky precision obtained during the UVIS transit observation of HAT-P-41\,b, we predicted that our observing strategy would result in a broadband eclipse precision of $\sim$35\,ppm per visit, increasing to $\sim$17\,ppm when stacking all four visits. However, our observations unfortunately yielded only a non-detection of the reflected light eclipse of LTT 9779\,b. Below, we lay out the factors that likely contributed to this result.

The first consideration is the amount of residual scatter in our observed light curves. In each visit, we reach an average residual light curve scatter of $\sim$200\,ppm, comparable to that obtained by \citet{wakeford_into_2020} for HAT-P-41\,b, despite their having collecting $\sim$30\% fewer photons due to the relative brightnesses of the HAT-P-41 and LTT 9779 host stars. We also note that this scatter does appear to be true white noise and not correlated noise that was not removed by the systematics models, as the residuals for each visit bin down exactly as would be expected for white noise. 

The UVIS simulator used in the proposal planning phase does not explicitly calculate the expected light curve scatter, only the spectroscopic precision. So we therefore constructed a grid of light curves with the same cadence and phase coverage as our observations, but different amounts of injected scatter, and fit them using the same methods as described in Section~\ref{sec: Light Curves HST} to back out the level of scatter which would enable us to obtain the predicted eclipse depth precision. After testing 50 realizations, we find that our observed scatter is $\sim$1.6 standard deviations higher than the average predicted scatter --- at the upper edge of consistency. 

The increased level of scatter does not appear to be, at root, an instrumental throughput issue. When comparing the median stellar spectrum we obtained with that predicted by the HST exposure time calculator \footnote{\url{https://etc.stsci.edu/etc/input/wfc3uvis/spectroscopic/}}, we find that the actual number of observed counts was within 5\% of the ETC predicted value when integrated over the entire UVIS waveband. There are some minor differences in the observed and ETC spectra, likely due to mismatches in the true properties of LTT 9779 and the available stellar templates. However, these differences average out when creating band-integrated white light curves. \citet{boehm_hustle_2025} also found that the specific treatment of the UVIS background can result in changes to the level of scatter in the white light curves. However, we only notice minor differences in our scatter when using a frame mode vs.\ median to remove the background.

Instead, the root of the increased level of scatter is likely the degrading sensitivity of the UVIS G280 optical components themselves. Using over a decade of archival observations of a flux standard star, \citet{alam2024} showed that the sensitivity of the UVIS G280 mode was decreasing by nearly half of a percent per year over the 2011--2024 period. \citet{boehm_hustle_2025} also found a lower-than-expected spectroscopic precision in their transmission spectrum. In all, we conclude that the instrument performed at the lower edge of our pre-observation expectations, and in light of this, recommend that UVIS G280 simulators, like the one used above, and the ETC be treated as best case scenarios for on-sky observations. 

The second factor is non-optimal coverage of the eclipse phases, which also likely contributed to the lower-than-anticipated observational precision. Due to scheduling constraints imposed by the comparable lengths of the planet's eclipse and the HST observability window at the latitude of LTT 9779 we did not obtain any coverage of the eclipse ingress or egress. Performing similar injection/recovery tests to those in Section~\ref{sec: Injection Recovery}, but assuming that two visits obtained partial coverage of the eclipse ingress and two of egress resulted in an upper limit roughly 10\% lower than the 113\,ppm we obtain from our actual observations.

\section{Conclusions} 
\label{sec: Conclusions}

In this study, we attempted the first spectroscopic study of an exoplanet eclipse using the HST/WFC3 UVIS G280 grism. With wavelength coverage from 0.2--0.8\,µm, the UVIS G280 grism offers the tantalizing possibility of detecting the scattering signatures from cloud condensates in the atmospheres of exoplanets --- complementary to cloud absorption signatures in the mid-infrared \citep[e.g.,][]{grant_jwst-tst_2023, dyrek_so2_2024, inglis_quartz_2024}. Such optical signatures could potentially directly identify the condensates responsible for the anomalously high albedos observed in the atmospheres of several hot, giant exoplanets \citep[e.g.,][]{shporer_atmospheric_2014, demory_inference_2013, hoyer_extremely_2023}. We targeted LTT 9779\,b, an ultra-hot-Neptune in the hot-Neptune desert, which has been previously be found to have a high geometric albedo from CHEOPS \citep{hoyer_extremely_2023} and JWST \citep{coulombe_highly_2025} optical observations.

However, we do not detect the eclipse of LTT 9779\,b either using a broadband white light curve, nor in any of our three 0.2\,µm-wide bins. Through injection-recovery tests, we set a 3-$\sigma$ upper limit of 113\,ppm on the 0.2--0.8\,µm broadband eclipse depth. We then investigate the reflective properties of LTT 9779\,b's dayside atmosphere by comparing our UVIS upper limit, along with previously published CHEOPS and TESS optical eclipse photometry to reflectance spectra generated using \texttt{VIRGA} and \texttt{PICASO}. In agreement with the analysis of \citet{coulombe_highly_2025} on a JWST NIRISS/SOSS emission spectrum of LTT 9779\,b, we find that the silicate species MgSiO$_3$ and Mg$_2$SiO$_4$ are the most likely candidates to explain the planet's high dayside albedo. Our models only offer weak constraints on cloud sedimentation efficiency or mixing strength. They do generally prefer small $f\rm _{sed}$ values, leading to smaller and more reflective condensate particles, and $K_{zz}$ values in line with expectations for ultra-hot atmospheres \citep[e.g.,][]{parmentier_3d_2013, komacek_vertical_2019} are not ruled out. 

We find that both non-optimal phase coverage of the eclipse ingress and egress as well as lower-than-expected observation precision, likely due to the degrading sensitivity of the UVIS spectroscopic modes \citep{alam2024}, contributed to our non-detection. It is therefore imperative for future UVIS programs account for these challenges in their observation planning. We also strongly advocate for robust testing for whether trends (astrophysical or otherwise) inferred in a time series observation are indeed statistically justified, particularly when attempting to push observational and instrumental limits.

\section*{Acknowledgements}
M.R.\ acknowledges funding from the Natural Sciences and Research Council of Canada (NSERC), the fonds de recherche du Québec --- Nature et Technologies (FRQNT), the centre de recherche en astrophysique du Québec (CRAQ), as well as support from the University of Bristol as a Visiting Research Student from 02-2023 to 05-2023. M.R.\ would also like to thank Abigael Radica and Lili Alderson for moral support. 
J.T.\ was supported by the Glasstone Benefaction, University of Oxford (Violette and Samuel Glasstone Research Fellowships in Science 2024). 
H.R.W.\ was funded by UK Research and Innovation (UKRI) framework under the UK government’s Horizon Europe funding guarantee for an ERC Starter Grant [grant number EP/Y006313/1].
J.S.J.\ greatfully acknowledges support by FONDECYT grant 1240738 and from the ANID BASAL project FB210003.
R.A.\ acknowledges the Swiss National Science Foundation (SNSF) support under the Post-Doc Mobility grant P500PT\_222212 and the support of the Institut Trottier de Recherche sur les Exoplanètes (iREx).
This project was undertaken with the financial support of the Canadian Space Agency.
This work is based on observations made with the NASA/ESA Hubble Space Telescope. 
This research has made use of the NASA Exoplanet Archive, which is operated by the California Institute of Technology, under contract with the National Aeronautics and Space Administration under the Exoplanet Exploration Program.

\section*{Data availability}
All data used in this study are publicly available from the Barbara A. Mikulski Archive for Space Telescopes\footnote{\url{https://mast.stsci.edu/portal/Mashup/Clients/Mast/Portal.html}}.

\section*{Software}
\texttt{astropy} \citep{astropy:2013, astropy:2018},
\texttt{exoTEDRF} \citep{radica_exotedrf_2024},
\texttt{exoUPRF} \citep{radica_exouprf_2024},
\texttt{ipython} \citep{PER-GRA:2007},
\texttt{juliet} \citep{espinoza_juliet_2019},
\texttt{lacosmic} \citep{vandokkum_cosmicray_2001},
\texttt{matplotlib} \citep{Hunter:2007},
\texttt{numpy} \citep{harris2020array},
\texttt{pymultinest} \citep{buchner_statistical_2016},
\texttt{scipy} \citep{2020SciPy-NMeth}

\bibliography{main}{}

\begin{thebibliography}{}
\makeatletter
\relax
\def\mn@urlcharsother{\let\do\@makeother \do\$\do\&\do\#\do\^\do\_\do\%\do\~}
\def\mn@doi{\begingroup\mn@urlcharsother \@ifnextchar [ {\mn@doi@} {\mn@doi@[]}}
\def\mn@doi@[#1]#2{\def\@tempa{#1}\ifx\@tempa\@empty \href {http://dx.doi.org/#2} {doi:#2}\else \href {http://dx.doi.org/#2} {#1}\fi \endgroup}
\def\mn@eprint#1#2{\mn@eprint@#1:#2::\@nil}
\def\mn@eprint@arXiv#1{\href {http://arxiv.org/abs/#1} {{\tt arXiv:#1}}}
\def\mn@eprint@dblp#1{\href {http://dblp.uni-trier.de/rec/bibtex/#1.xml} {dblp:#1}}
\def\mn@eprint@#1:#2:#3:#4\@nil{\def\@tempa {#1}\def\@tempb {#2}\def\@tempc {#3}\ifx \@tempc \@empty \let \@tempc \@tempb \let \@tempb \@tempa \fi \ifx \@tempb \@empty \def\@tempb {arXiv}\fi \@ifundefined {mn@eprint@\@tempb}{\@tempb:\@tempc}{\expandafter \expandafter \csname mn@eprint@\@tempb\endcsname \expandafter{\@tempc}}}

\bibitem[\protect\citeauthoryear{Ackerman \& Marley}{Ackerman \& Marley}{2001}]{ackerman_precipitating_2001}
Ackerman A.~S.,  Marley M.~S.,  2001, \mn@doi [ApJ] {10.1086/321540}, 556, 872

\bibitem[\protect\citeauthoryear{{Alam}, Annalisa, Benjamin, Debopam  \& Sylvia}{{Alam} et~al.}{2024}]{alam2024}
{Alam} M.,  Annalisa C.,  Benjamin K.,  Debopam S.,   Sylvia B.,  2024, {Sensitivity Evolution of the HST WFC3/UVIS G280 Grism}, Instrument Science Report WFC3 2024-11, 18 pages

\bibitem[\protect\citeauthoryear{Albert et~al.,}{Albert et~al.}{2023}]{albert_near_2023}
Albert L.,  et~al., 2023, \mn@doi [PASP] {10.1088/1538-3873/acd7a3}, 135, 075001

\bibitem[\protect\citeauthoryear{Anderson \& Darling}{Anderson \& Darling}{1952}]{anderson_asymptotic_1952}
Anderson T.~W.,  Darling D.~A.,  1952, The annals of mathematical statistics, pp 193--212

\bibitem[\protect\citeauthoryear{Angerhausen, DeLarme  \& Morse}{Angerhausen et~al.}{2015}]{angerhausen_comprehensive_2015}
Angerhausen D.,  DeLarme E.,   Morse J.~A.,  2015, \mn@doi [Publications of the Astronomical Society of the Pacific] {10.1086/683797}, 127, 1113

\bibitem[\protect\citeauthoryear{Arcangeli et~al.,}{Arcangeli et~al.}{2018}]{arcangeli_h_2018}
Arcangeli J.,  et~al., 2018, \mn@doi [ApJ] {10.3847/2041-8213/aab272}, 855, L30

\bibitem[\protect\citeauthoryear{{Astropy Collaboration} et~al.,}{{Astropy Collaboration} et~al.}{2013}]{astropy:2013}
{Astropy Collaboration} et~al., 2013, \mn@doi [\aap] {10.1051/0004-6361/201322068}, \href {http://adsabs.harvard.edu/abs/2013A%26A...558A..33A} {558, A33}

\bibitem[\protect\citeauthoryear{{Astropy Collaboration} et~al.,}{{Astropy Collaboration} et~al.}{2018}]{astropy:2018}
{Astropy Collaboration} et~al., 2018, \mn@doi [\aj] {10.3847/1538-3881/aabc4f}, \href {https://ui.adsabs.harvard.edu/abs/2018AJ....156..123A} {156, 123}

\bibitem[\protect\citeauthoryear{{Batalha}, {Marley}, {Lewis}  \& {Fortney}}{{Batalha} et~al.}{2019}]{batalha_exoplanet_2019}
{Batalha} N.~E.,  {Marley} M.~S.,  {Lewis} N.~K.,   {Fortney} J.~J.,  2019, \mn@doi [\apj] {10.3847/1538-4357/ab1b51}, \href {https://ui.adsabs.harvard.edu/abs/2019ApJ...878...70B} {878, 70}

\bibitem[\protect\citeauthoryear{Baxter et~al.,}{Baxter et~al.}{2020}]{baxter_transition_2020}
Baxter C.,  et~al., 2020, \mn@doi [A\&A] {10.1051/0004-6361/201937394}, 639, A36

\bibitem[\protect\citeauthoryear{Beatty, Madhusudhan, Tsiaras, Zhao, Gilliland, Knutson, Shporer  \& Wright}{Beatty et~al.}{2017}]{beatty_evidence_2017}
Beatty T.~G.,  Madhusudhan N.,  Tsiaras A.,  Zhao M.,  Gilliland R.~L.,  Knutson H.~A.,  Shporer A.,   Wright J.~T.,  2017, \mn@doi [AJ] {10.3847/1538-3881/aa899b}, 154, 158

\bibitem[\protect\citeauthoryear{Bell et~al.,}{Bell et~al.}{2017}]{bell_very_2017}
Bell T.~J.,  et~al., 2017, \mn@doi [ApJ] {10.3847/2041-8213/aa876c}, 847, L2

\bibitem[\protect\citeauthoryear{Benneke \& Seager}{Benneke \& Seager}{2013}]{benneke_how_2013}
Benneke B.,  Seager S.,  2013, \mn@doi [ApJ] {10.1088/0004-637X/778/2/153}, 778, 153

\bibitem[\protect\citeauthoryear{Boehm et~al.,}{Boehm et~al.}{2025}]{boehm_hustle_2025}
Boehm V.~A.,  et~al., 2025, \mn@doi [AJ] {10.3847/1538-3881/ad8dde}, 169, 23

\bibitem[\protect\citeauthoryear{Brandeker et~al.,}{Brandeker et~al.}{2022}]{brandeker_cheops_2022}
Brandeker A.,  et~al., 2022, \mn@doi [A\&A] {10.1051/0004-6361/202243082}, 659, L4

\bibitem[\protect\citeauthoryear{Buchner}{Buchner}{2016}]{buchner_statistical_2016}
Buchner J.,  2016, \mn@doi [Stat Comput] {10.1007/s11222-014-9512-y}, 26, 383

\bibitem[\protect\citeauthoryear{Changeat et~al.,}{Changeat et~al.}{2022}]{changeat_five_2022}
Changeat Q.,  et~al., 2022, \mn@doi [ApJS] {10.3847/1538-4365/ac5cc2}, 260, 3

\bibitem[\protect\citeauthoryear{Charbonneau et~al.,}{Charbonneau et~al.}{2005}]{charbonneau_detection_2005}
Charbonneau D.,  et~al., 2005, \mn@doi [ApJ] {10.1086/429991}, 626, 523

\bibitem[\protect\citeauthoryear{Chubb, Samra, Helling, Carone  \& Stam}{Chubb et~al.}{2024}]{chubb_dark_2024}
Chubb K.~L.,  Samra D.,  Helling C.,  Carone L.,   Stam D.~M.,  2024, \mn@doi [Monthly Notices of the Royal Astronomical Society] {10.1093/mnras/stae1916}, 533, 1503

\bibitem[\protect\citeauthoryear{Coulombe et~al.,}{Coulombe et~al.}{2023}]{coulombe_broadband_2023}
Coulombe L.-P.,  et~al., 2023, \mn@doi [Nature] {10.1038/s41586-023-06230-1}, 620, 292

\bibitem[\protect\citeauthoryear{{Coulombe} et~al.,}{{Coulombe} et~al.}{2025}]{coulombe_highly_2025}
{Coulombe} L.-P.,  et~al., 2025, \mn@doi [arXiv e-prints] {10.48550/arXiv.2501.14016}, \href {https://ui.adsabs.harvard.edu/abs/2025arXiv250114016C} {p. arXiv:2501.14016}

\bibitem[\protect\citeauthoryear{Crossfield et~al.,}{Crossfield et~al.}{2020}]{crossfield_phase_2020}
Crossfield I. J.~M.,  et~al., 2020, \mn@doi [ApJ] {10.3847/2041-8213/abbc71}, 903, L7

\bibitem[\protect\citeauthoryear{Deming, Seager, Richardson  \& Harrington}{Deming et~al.}{2005}]{deming_infrared_2005}
Deming D.,  Seager S.,  Richardson L.~J.,   Harrington J.,  2005, \mn@doi [Nature] {10.1038/nature03507}, 434, 740

\bibitem[\protect\citeauthoryear{Deming et~al.,}{Deming et~al.}{2015}]{deming_spitzer_2015}
Deming D.,  et~al., 2015, \mn@doi [ApJ] {10.1088/0004-637X/805/2/132}, 805, 132

\bibitem[\protect\citeauthoryear{Demory et~al.,}{Demory et~al.}{2013}]{demory_inference_2013}
Demory B.-O.,  et~al., 2013, \mn@doi [ApJ] {10.1088/2041-8205/776/2/L25}, 776, L25

\bibitem[\protect\citeauthoryear{Doyon et~al.,}{Doyon et~al.}{2023}]{doyon_near_2023}
Doyon R.,  et~al., 2023, \mn@doi [PASP] {10.1088/1538-3873/acd41b}, 135, 098001

\bibitem[\protect\citeauthoryear{Dragomir et~al.,}{Dragomir et~al.}{2020}]{dragomir_spitzer_2020}
Dragomir D.,  et~al., 2020, \mn@doi [ApJ] {10.3847/2041-8213/abbc70}, 903, L6

\bibitem[\protect\citeauthoryear{Dyrek et~al.,}{Dyrek et~al.}{2024}]{dyrek_so2_2024}
Dyrek A.,  et~al., 2024, \mn@doi [Nature] {10.1038/s41586-023-06849-0}, 625, 51

\bibitem[\protect\citeauthoryear{Edwards et~al.,}{Edwards et~al.}{2023}]{edwards_characterizing_2023}
Edwards B.,  et~al., 2023, \mn@doi [AJ] {10.3847/1538-3881/acea77}, 166, 158

\bibitem[\protect\citeauthoryear{Espinoza, Kossakowski  \& Brahm}{Espinoza et~al.}{2019}]{espinoza_juliet_2019}
Espinoza N.,  Kossakowski D.,   Brahm R.,  2019, \mn@doi [Monthly Notices of the Royal Astronomical Society] {10.1093/mnras/stz2688}, 490, 2262

\bibitem[\protect\citeauthoryear{{Esteves}, {De Mooij}  \& {Jayawardhana}}{{Esteves} et~al.}{2015}]{esteves_changing_2015}
{Esteves} L.~J.,  {De Mooij} E. J.~W.,   {Jayawardhana} R.,  2015, \mn@doi [\apj] {10.1088/0004-637X/804/2/150}, \href {https://ui.adsabs.harvard.edu/abs/2015ApJ...804..150E} {804, 150}

\bibitem[\protect\citeauthoryear{Evans et~al.,}{Evans et~al.}{2013}]{evans_deep_2013}
Evans T.~M.,  et~al., 2013, \mn@doi [ApJ] {10.1088/2041-8205/772/2/L16}, 772, L16

\bibitem[\protect\citeauthoryear{Evans et~al.,}{Evans et~al.}{2017}]{evans_ultrahot_2017}
Evans T.~M.,  et~al., 2017, \mn@doi [Nature] {10.1038/nature23266}, 548, 58

\bibitem[\protect\citeauthoryear{Feinstein et~al.,}{Feinstein et~al.}{2023}]{feinstein_early_2023}
Feinstein A.~D.,  et~al., 2023, \mn@doi [Nature] {10.1038/s41586-022-05674-1}, 614, 670

\bibitem[\protect\citeauthoryear{Fraine et~al.,}{Fraine et~al.}{2021}]{fraine_dark_2021}
Fraine J.,  et~al., 2021, \mn@doi [AJ] {10.3847/1538-3881/abe8d6}, 161, 269

\bibitem[\protect\citeauthoryear{{Gao}, {Marley}  \& {Ackerman}}{{Gao} et~al.}{2018}]{gao_sedimentation_2018}
{Gao} P.,  {Marley} M.~S.,   {Ackerman} A.~S.,  2018, \mn@doi [\apj] {10.3847/1538-4357/aab0a1}, \href {https://ui.adsabs.harvard.edu/abs/2018ApJ...855...86G} {855, 86}

\bibitem[\protect\citeauthoryear{Gao et~al.,}{Gao et~al.}{2020}]{gao_aerosol_2020}
Gao P.,  et~al., 2020, \mn@doi [Nat Astron] {10.1038/s41550-020-1114-3}, 4, 951

\bibitem[\protect\citeauthoryear{Gao, Wakeford, Moran  \& Parmentier}{Gao et~al.}{2021}]{gao_aerosols_2021}
Gao P.,  Wakeford H.~R.,  Moran S.~E.,   Parmentier V.,  2021, \mn@doi [JGR Planets] {10.1029/2020JE006655}, 126, e2020JE006655

\bibitem[\protect\citeauthoryear{Grant et~al.,}{Grant et~al.}{2023}]{grant_jwst-tst_2023}
Grant D.,  et~al., 2023, \mn@doi [ApJL] {10.3847/2041-8213/acfc3b}, 956, L29

\bibitem[\protect\citeauthoryear{Harris et~al.,}{Harris et~al.}{2020}]{harris2020array}
Harris C.~R.,  et~al., 2020, \mn@doi [Nature] {10.1038/s41586-020-2649-2}, 585, 357

\bibitem[\protect\citeauthoryear{Heng \& Demory}{Heng \& Demory}{2013}]{heng_understanding_2013}
Heng K.,  Demory B.-O.,  2013, \mn@doi [ApJ] {10.1088/0004-637X/777/2/100}, 777, 100

\bibitem[\protect\citeauthoryear{Hoyer et~al.,}{Hoyer et~al.}{2023}]{hoyer_extremely_2023}
Hoyer S.,  et~al., 2023, \mn@doi [A\&A] {10.1051/0004-6361/202346117}, 675, A81

\bibitem[\protect\citeauthoryear{Hubeny, Burrows  \& Sudarsky}{Hubeny et~al.}{2003}]{hubeny_possible_2003}
Hubeny I.,  Burrows A.,   Sudarsky D.,  2003, \mn@doi [ApJ] {10.1086/377080}, 594, 1011

\bibitem[\protect\citeauthoryear{Hunter}{Hunter}{2007}]{Hunter:2007}
Hunter J.~D.,  2007, \mn@doi [Computing in Science \& Engineering] {10.1109/MCSE.2007.55}, 9, 90

\bibitem[\protect\citeauthoryear{Husser, Wende-von Berg, Dreizler, Homeier, Reiners, Barman  \& Hauschildt}{Husser et~al.}{2013}]{husser_new_2013}
Husser T.-O.,  Wende-von Berg S.,  Dreizler S.,  Homeier D.,  Reiners A.,  Barman T.,   Hauschildt P.~H.,  2013, \mn@doi [A\&A] {10.1051/0004-6361/201219058}, 553, A6

\bibitem[\protect\citeauthoryear{Inglis et~al.,}{Inglis et~al.}{2024}]{inglis_quartz_2024}
Inglis J.,  et~al., 2024, \mn@doi [ApJL] {10.3847/2041-8213/ad725e}, 973, L41

\bibitem[\protect\citeauthoryear{Jenkins et~al.,}{Jenkins et~al.}{2020}]{jenkins_ultrahot_2020}
Jenkins J.~S.,  et~al., 2020, \mn@doi [Nat Astron] {10.1038/s41550-020-1142-z}, 4, 1148

\bibitem[\protect\citeauthoryear{Kirk, Wheatley, Louden, Doyle, Skillen, McCormac, Irwin  \& Karjalainen}{Kirk et~al.}{2017}]{kirk_rayleigh_2017}
Kirk J.,  Wheatley P.~J.,  Louden T.,  Doyle A.~P.,  Skillen I.,  McCormac J.,  Irwin P. G.~J.,   Karjalainen R.,  2017, \mn@doi [Monthly Notices of the Royal Astronomical Society] {10.1093/mnras/stx752}, 468, 3907

\bibitem[\protect\citeauthoryear{Komacek, Showman  \& Parmentier}{Komacek et~al.}{2019}]{komacek_vertical_2019}
Komacek T.~D.,  Showman A.~P.,   Parmentier V.,  2019, \mn@doi [ApJ] {10.3847/1538-4357/ab338b}, 881, 152

\bibitem[\protect\citeauthoryear{Kreidberg}{Kreidberg}{2015}]{kreidberg_batman_2015}
Kreidberg L.,  2015, \mn@doi [Publications of the Astronomical Society of the Pacific] {10.1086/683602}, 127, 1161

\bibitem[\protect\citeauthoryear{Krenn et~al.,}{Krenn et~al.}{2023}]{krenn_geometric_2023}
Krenn A.~F.,  et~al., 2023, \mn@doi [A\&A] {10.1051/0004-6361/202245016}, 672, A24

\bibitem[\protect\citeauthoryear{Lewis et~al.,}{Lewis et~al.}{2020}]{lewis_into_2020}
Lewis N.~K.,  et~al., 2020, \mn@doi [ApJL] {10.3847/2041-8213/abb77f}, 902, L19

\bibitem[\protect\citeauthoryear{{Lothringer} et~al.,}{{Lothringer} et~al.}{2022}]{lothringer_uv_2022}
{Lothringer} J.~D.,  et~al., 2022, \mn@doi [\nat] {10.1038/s41586-022-04453-2}, \href {https://ui.adsabs.harvard.edu/abs/2022Natur.604...49L} {604, 49}

\bibitem[\protect\citeauthoryear{Mansfield et~al.,}{Mansfield et~al.}{2021}]{mansfield_unique_2021}
Mansfield M.,  et~al., 2021, \mn@doi [Nat Astron] {10.1038/s41550-021-01455-4}, 5, 1224

\bibitem[\protect\citeauthoryear{Massey~Jr}{Massey~Jr}{1951}]{massey_kolmogorov_1951}
Massey~Jr F.~J.,  1951, Journal of the American statistical Association, 46, 68

\bibitem[\protect\citeauthoryear{Mazeh, Holczer  \& Faigler}{Mazeh et~al.}{2016}]{mazeh_dearth_2016}
Mazeh T.,  Holczer T.,   Faigler S.,  2016, \mn@doi [A\&A] {10.1051/0004-6361/201528065}, 589, A75

\bibitem[\protect\citeauthoryear{Parmentier, Showman  \& Lian}{Parmentier et~al.}{2013}]{parmentier_3d_2013}
Parmentier V.,  Showman A.~P.,   Lian Y.,  2013, \mn@doi [A\&A] {10.1051/0004-6361/201321132}, 558, A91

\bibitem[\protect\citeauthoryear{Parmentier, Fortney, Showman, Morley  \& Marley}{Parmentier et~al.}{2016}]{parmentier_transitions_2016}
Parmentier V.,  Fortney J.~J.,  Showman A.~P.,  Morley C.,   Marley M.~S.,  2016, \mn@doi [ApJ] {10.3847/0004-637X/828/1/22}, 828, 22

\bibitem[\protect\citeauthoryear{Parmentier et~al.,}{Parmentier et~al.}{2018}]{parmentier_thermal_2018}
Parmentier V.,  et~al., 2018, \mn@doi [A\&A] {10.1051/0004-6361/201833059}, 617, A110

\bibitem[\protect\citeauthoryear{Parmentier, Showman  \& Fortney}{Parmentier et~al.}{2020}]{parmentier_cloudy_2020}
Parmentier V.,  Showman A.~P.,   Fortney J.~J.,  2020, \mn@doi [Monthly Notices of the Royal Astronomical Society] {10.1093/mnras/staa3418}, 501, 78

\bibitem[\protect\citeauthoryear{P\'erez \& Granger}{P\'erez \& Granger}{2007}]{PER-GRA:2007}
P\'erez F.,  Granger B.~E.,  2007, \mn@doi [Computing in Science and Engineering] {10.1109/MCSE.2007.53}, 9, 21

\bibitem[\protect\citeauthoryear{{Pirzkal}}{{Pirzkal}}{2020}]{pirzkal2020}
{Pirzkal} N.,  2020, {Updated Calibration of the UVIS G280 Grism}, Instrument Science Report WFC3 2020-9, 27 pages

\bibitem[\protect\citeauthoryear{{Pirzkal}, {Hilbert}  \& {Rothberg}}{{Pirzkal} et~al.}{2017}]{pirzkal2017}
{Pirzkal} N.,  {Hilbert} B.,   {Rothberg} B.,  2017, {Trace and Wavelength Calibrations of the UVIS G280 +1/-1 Grism Orders}, Instrument Science Report WFC3 2017-20, 15 pages

\bibitem[\protect\citeauthoryear{Pont, Knutson, Gilliland, Moutou  \& Charbonneau}{Pont et~al.}{2008}]{pont_detection_2008}
Pont F.,  Knutson H.,  Gilliland R.~L.,  Moutou C.,   Charbonneau D.,  2008, \mn@doi [Monthly Notices of the Royal Astronomical Society] {10.1111/j.1365-2966.2008.12852.x}, 385, 109

\bibitem[\protect\citeauthoryear{{Radica}}{{Radica}}{2024a}]{radica_exouprf_2024}
{Radica} M.,  2024a, {radicamc/exoUPRF: v1.0.1}, \mn@doi{10.5281/zenodo.12628066}

\bibitem[\protect\citeauthoryear{Radica}{Radica}{2024b}]{radica_exotedrf_2024}
Radica M.,  2024b, \mn@doi [JOSS] {10.21105/joss.06898}, 9, 6898

\bibitem[\protect\citeauthoryear{Radica et~al.,}{Radica et~al.}{2022}]{radica_applesoss_2022}
Radica M.,  et~al., 2022, \mn@doi [PASP] {10.1088/1538-3873/ac9430}, 134, 104502

\bibitem[\protect\citeauthoryear{Radica et~al.,}{Radica et~al.}{2023}]{radica_awesome_2023}
Radica M.,  et~al., 2023, \mn@doi [Monthly Notices of the Royal Astronomical Society] {10.1093/mnras/stad1762}, 524, 835

\bibitem[\protect\citeauthoryear{Radica et~al.,}{Radica et~al.}{2024}]{radica_muted_2024}
Radica M.,  et~al., 2024, \mn@doi [ApJL] {10.3847/2041-8213/ad20e4}, 962, L20

\bibitem[\protect\citeauthoryear{Radica et~al.,}{Radica et~al.}{2025}]{radica_promise_2025}
Radica M.,  et~al., 2025, \mn@doi [ApJL] {10.3847/2041-8213/ada381}, 979, L5

\bibitem[\protect\citeauthoryear{Reyes et~al.,}{Reyes et~al.}{2025}]{reyes_closer_2025}
Reyes R.~R.,  et~al., 2025, \mn@doi [A\&A] {10.1051/0004-6361/202451044}

\bibitem[\protect\citeauthoryear{{Rooney}, {Batalha}, {Gao}  \& {Marley}}{{Rooney} et~al.}{2022}]{rooney_new_2022}
{Rooney} C.~M.,  {Batalha} N.~E.,  {Gao} P.,   {Marley} M.~S.,  2022, \mn@doi [\apj] {10.3847/1538-4357/ac307a}, \href {https://ui.adsabs.harvard.edu/abs/2022ApJ...925...33R} {925, 33}

\bibitem[\protect\citeauthoryear{Schwartz \& Cowan}{Schwartz \& Cowan}{2015}]{schwartz_balancing_2015}
Schwartz J.~C.,  Cowan N.~B.,  2015, \mn@doi [Monthly Notices of the Royal Astronomical Society] {10.1093/mnras/stv470}, 449, 4192

\bibitem[\protect\citeauthoryear{Seager \& Mallen‐Ornelas}{Seager \& Mallen‐Ornelas}{2003}]{seager_unique_2003}
Seager S.,  Mallen‐Ornelas G.,  2003, \mn@doi [ApJ] {10.1086/346105}, 585, 1038

\bibitem[\protect\citeauthoryear{Seager, Whitney  \& Sasselov}{Seager et~al.}{2000}]{seager_photometric_2000}
Seager S.,  Whitney B.~A.,   Sasselov D.~D.,  2000, \mn@doi [ApJ] {10.1086/309292}, 540, 504

\bibitem[\protect\citeauthoryear{Seager, Richardson, Hansen, Menou, Cho  \& Deming}{Seager et~al.}{2005}]{seager_dayside_2005}
Seager S.,  Richardson L.~J.,  Hansen B. M.~S.,  Menou K.,  Cho J.~Y.,   Deming D.,  2005, \mn@doi [ApJ] {10.1086/444411}, 632, 1122

\bibitem[\protect\citeauthoryear{{Shporer} et~al.,}{{Shporer} et~al.}{2014}]{shporer_atmospheric_2014}
{Shporer} A.,  et~al., 2014, \mn@doi [\apj] {10.1088/0004-637X/788/1/92}, \href {https://ui.adsabs.harvard.edu/abs/2014ApJ...788...92S} {788, 92}

\bibitem[\protect\citeauthoryear{Sing et~al.,}{Sing et~al.}{2016}]{sing_continuum_2016}
Sing D.~K.,  et~al., 2016, \mn@doi [Nature] {10.1038/nature16068}, 529, 59

\bibitem[\protect\citeauthoryear{Sing et~al.,}{Sing et~al.}{2019}]{sing_hubble_2019}
Sing D.~K.,  et~al., 2019, \mn@doi [AJ] {10.3847/1538-3881/ab2986}, 158, 91

\bibitem[\protect\citeauthoryear{{Singh} et~al.,}{{Singh} et~al.}{2024}]{singh_cheops_2024}
{Singh} V.,  et~al., 2024, \mn@doi [\aap] {10.1051/0004-6361/202347533}, \href {https://ui.adsabs.harvard.edu/abs/2024A&A...683A...1S} {683, A1}

\bibitem[\protect\citeauthoryear{Spiegel, Silverio  \& Burrows}{Spiegel et~al.}{2009}]{spiegel_can_2009}
Spiegel D.~S.,  Silverio K.,   Burrows A.,  2009, \mn@doi [ApJ] {10.1088/0004-637X/699/2/1487}, 699, 1487

\bibitem[\protect\citeauthoryear{Sudarsky, Burrows  \& Pinto}{Sudarsky et~al.}{2000}]{sudarsky_albedo_2000}
Sudarsky D.,  Burrows A.,   Pinto P.,  2000, \mn@doi [ApJ] {10.1086/309160}, 538, 885

\bibitem[\protect\citeauthoryear{Szabó \& Kiss}{Szabó \& Kiss}{2011}]{szabo_short-period_2011}
Szabó G.~M.,  Kiss L.~L.,  2011, \mn@doi [ApJ] {10.1088/2041-8205/727/2/L44}, 727, L44

\bibitem[\protect\citeauthoryear{Taylor \& Parmentier}{Taylor \& Parmentier}{2023}]{taylor_another_2023}
Taylor J.,  Parmentier V.,  2023, \mn@doi [Monthly Notices of the Royal Astronomical Society] {10.1093/mnras/stad2287}, 526, 2133

\bibitem[\protect\citeauthoryear{Taylor, Parmentier, Line, Lee, Irwin  \& Aigrain}{Taylor et~al.}{2021}]{taylor_how_2021}
Taylor J.,  Parmentier V.,  Line M.~R.,  Lee E. K.~H.,  Irwin P. G.~J.,   Aigrain S.,  2021, mnras, 506, 24

\bibitem[\protect\citeauthoryear{Taylor et~al.,}{Taylor et~al.}{2023}]{taylor_awesome_2023}
Taylor J.,  et~al., 2023, \mn@doi [Monthly Notices of the Royal Astronomical Society] {10.1093/mnras/stad1547}, 524, 817

\bibitem[\protect\citeauthoryear{Thorngren, Fortney, Murray-Clay  \& Lopez}{Thorngren et~al.}{2016}]{thorngren_massmetallicity_2016}
Thorngren D.~P.,  Fortney J.~J.,  Murray-Clay R.~A.,   Lopez E.~D.,  2016, \mn@doi [ApJ] {10.3847/0004-637X/831/1/64}, 831, 64

\bibitem[\protect\citeauthoryear{Virtanen et~al.,}{Virtanen et~al.}{2020}]{2020SciPy-NMeth}
Virtanen P.,  et~al., 2020, \mn@doi [Nature Methods] {10.1038/s41592-019-0686-2}, \href {https://rdcu.be/b08Wh} {17, 261}

\bibitem[\protect\citeauthoryear{Vissapragada et~al.,}{Vissapragada et~al.}{2024}]{vissapragada_high-resolution_2024}
Vissapragada S.,  et~al., 2024, \mn@doi [ApJL] {10.3847/2041-8213/ad23cf}, 962, L19

\bibitem[\protect\citeauthoryear{Visscher, Lodders  \& Fegley}{Visscher et~al.}{2010}]{visscher_atmospheric_2010}
Visscher C.,  Lodders K.,   Fegley B.,  2010, \mn@doi [ApJ] {10.1088/0004-637X/716/2/1060}, 716, 1060

\bibitem[\protect\citeauthoryear{Wakeford \& Sing}{Wakeford \& Sing}{2015}]{wakeford_transmission_2015}
Wakeford H.~R.,  Sing D.~K.,  2015, \mn@doi [A\&A] {10.1051/0004-6361/201424207}, 573, A122

\bibitem[\protect\citeauthoryear{{Wakeford}, {Sing}, {Evans}, {Deming}  \& {Mandell}}{{Wakeford} et~al.}{2016}]{wakeford_marginalizing_2016}
{Wakeford} H.~R.,  {Sing} D.~K.,  {Evans} T.,  {Deming} D.,   {Mandell} A.,  2016, \mn@doi [\apj] {10.3847/0004-637X/819/1/10}, \href {https://ui.adsabs.harvard.edu/abs/2016ApJ...819...10W} {819, 10}

\bibitem[\protect\citeauthoryear{Wakeford, Visscher, Lewis, Kataria, Marley, Fortney  \& Mandell}{Wakeford et~al.}{2017}]{wakeford_high-temperature_2017}
Wakeford H.~R.,  Visscher C.,  Lewis N.~K.,  Kataria T.,  Marley M.~S.,  Fortney J.~J.,   Mandell A.~M.,  2017, \mn@doi [Mon. Not. R. Astron. Soc.] {10.1093/mnras/stw2639}, 464, 4247

\bibitem[\protect\citeauthoryear{Wakeford et~al.,}{Wakeford et~al.}{2020}]{wakeford_into_2020}
Wakeford H.~R.,  et~al., 2020, \mn@doi [AJ] {10.3847/1538-3881/ab7b78}, 159, 204

\bibitem[\protect\citeauthoryear{Welbanks, Madhusudhan, Allard, Hubeny, Spiegelman  \& Leininger}{Welbanks et~al.}{2019}]{welbanks_massmetallicity_2019}
Welbanks L.,  Madhusudhan N.,  Allard N.~F.,  Hubeny I.,  Spiegelman F.,   Leininger T.,  2019, \mn@doi [ApJ] {10.3847/2041-8213/ab5a89}, 887, L20

\bibitem[\protect\citeauthoryear{{Wong} et~al.,}{{Wong} et~al.}{2021}]{Wong2021}
{Wong} I.,  et~al., 2021, \mn@doi [\aj] {10.3847/1538-3881/ac0c7d}, \href {https://ui.adsabs.harvard.edu/abs/2021AJ....162..127W} {162, 127}

\bibitem[\protect\citeauthoryear{van Dokkum}{van Dokkum}{2001}]{vandokkum_cosmicray_2001}
van Dokkum P.,  2001, \mn@doi [PUBL ASTRON SOC PAC] {10.1086/323894}, 113, 1420

\makeatother
\end{thebibliography}
\bibliographystyle{mnras}

\appendix
\section{Additional Figures}

\begin{figure*} 
	\centering
	\includegraphics[width=0.95\textwidth]{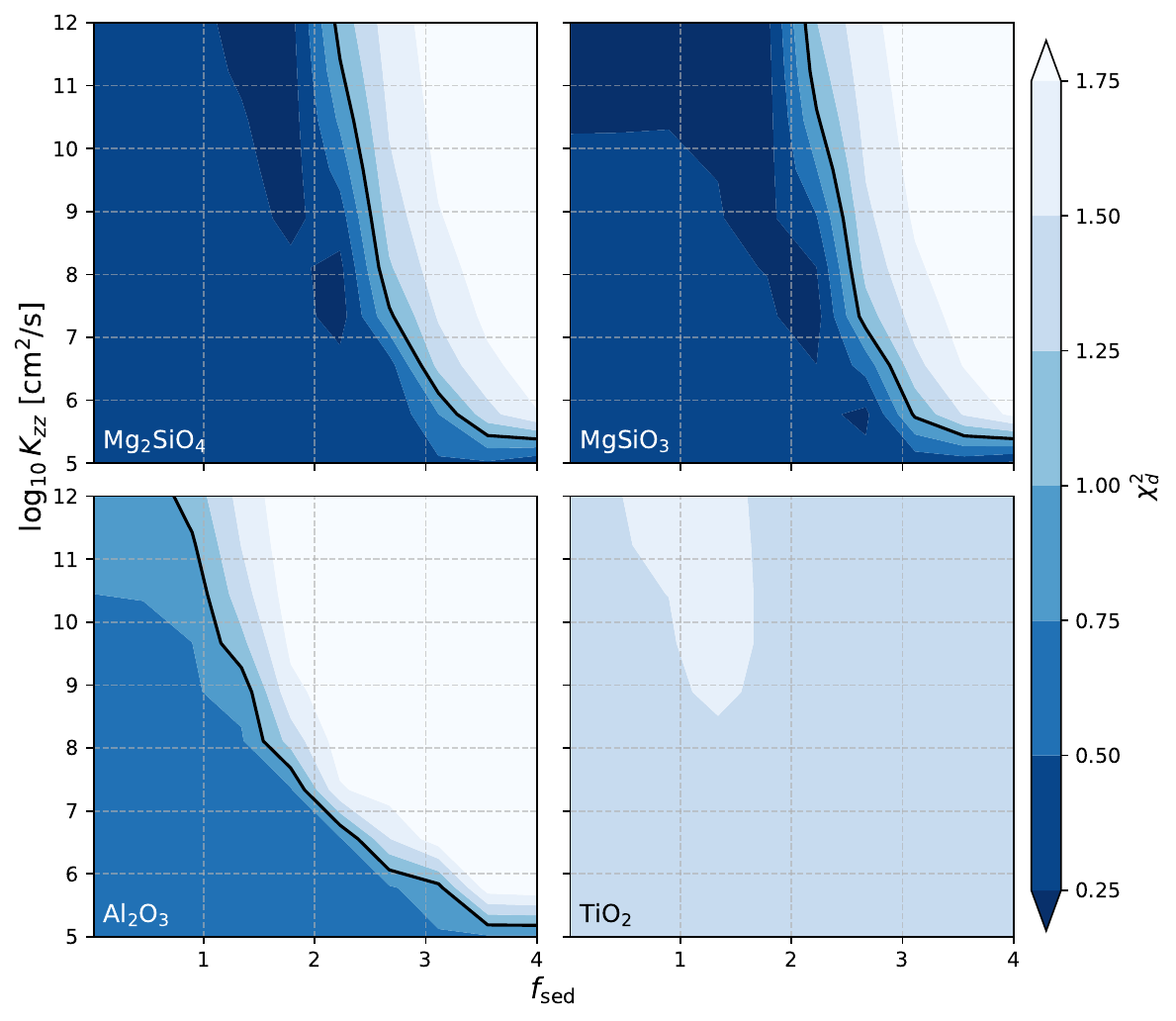}
    \caption{Same as Figure~\ref{fig: Chi2 Maps} but when excluding the CHEOPS eclipse depth from the analysis. The general trends in $f\rm _{sed}$ and $K_{zz}$ from Figure~\ref{fig: Chi2 Maps} remain qualitatively unchanged. The solid black lines denote curves of $\chi^2_d$=1. 
    \label{fig: Chi2 Map No CHEOPS}}
\end{figure*}

Figure~\ref{fig: Chi2 Map No CHEOPS} shows the same $\chi^2_d$ maps as Figure~\ref{fig: Chi2 Maps}, but when we instead exclude the CHEOPS eclipse depth from the analysis. 

\bsp	
\label{lastpage}
\end{document}